\documentclass[letter]{aa} % for the letters 
\usepackage{graphicx}
\usepackage{cprotect}
\usepackage{gensymb} 
\def\ms{\hbox{\,m\,s$^{-1}$}}         %m.s -1
\def\cms{\hbox{\,cm\,s$^{-1}$}}       %cm.s -1
\def\m2s2{\hbox{\,m$^{2}$\,s$^{-2}$}} %m2.s -2
\def\kms{\hbox{\,km\,s$^{-1}$}}       %km.s -1
\def\Msun{\hbox{$M_{\odot}$}}             %Msun
\def\Rsun{\hbox{$R_{\odot}$}}
\def\Mjup{\hbox{$\mathrm{M}_{\rm J}$}}

\def\Kepler{{\it Kepler}}
\def\ktwo{{\it K2}}
\def\ten[#1]{$\;\times 10^{#1}$}

\def\logg{$\log g$}

\def\epic{\mbox{K2-3}}
\def\solar {\ifmmode_{\mathord\odot} \else $_{\mathord\odot}$ \fi}% _solar
\def\jup {\ifmmode_{\mathrm{Jup}} \else $_{\mathrm{Jup}}$ \fi}% _jup
\def\earth {\ifmmode_{\mathord\oplus} \else $_{\mathord\oplus}$ \fi}% _earth
\def\Msol {\ifmmode {\,\mathrm{M}\solar} \else \,M\solar \fi}     % solar mass
\def\Rsol {\ifmmode {\,\mathrm{R}\solar} \else R\solar \fi}     % solar radius
\def\Lsol {\ifmmode {\,\mathrm{L}\solar} \else L\solar \fi}     % solar radius
\def\Mjup {\ifmmode {\,\mathrm{M}\jup} \else M\jup \fi}
\def\Mearth {\ifmmode {\,\mathrm{M}\earth} \else M\earth \fi}
\def\Rearth {\ifmmode {\,\mathrm{R}\earth} \else R\earth \fi}

\def\mps {\ifmmode {\,\mathrm{m\,s^{-1}}} \else $\mathrm{m\,s^{-1}}$ \fi}     % meter per sec

\begin{document}

   \title{A HARPS view on \epic\thanks{Based on observations made with the HARPS instrument on the ESO 3.6 m telescope under the program ID 191-C0873 at Cerro La Silla (Chile).}}

\author{J.M.~Almenara\inst{\ref{grenoble1},\ref{grenoble2}} 
\and N.~Astudillo-Defru\inst{\ref{grenoble1},\ref{grenoble2}}
\and X.~Bonfils\inst{\ref{grenoble1},\ref{grenoble2}}
\and T.~Forveille\inst{\ref{grenoble1},\ref{grenoble2}}
\and A.~Santerne\inst{\ref{porto2}}
\and S.~Albrecht\inst{\ref{aarhus}}
\and S.C.C.~Barros\inst{\ref{lam}}
\and F.~Bouchy\inst{\ref{lam}}
\and X.~Delfosse\inst{\ref{grenoble1},\ref{grenoble2}}
\and O.~Demangeon\inst{\ref{lam}}
\and R.F.~D\'{i}az\inst{\ref{geneva}}
\and G.~H\'ebrard\inst{\ref{ohp},\ref{iap}}
\and M.~Mayor\inst{\ref{geneva}}
\and V.~Neves\inst{\ref{natal}}
\and P.~Rojo\inst{\ref{santiago}}
\and N.C.~Santos\inst{\ref{porto2},\ref{porto3}}
\and A.~W\"unsche\inst{\ref{grenoble1},\ref{grenoble2}}
             }

\institute{
Univ. Grenoble Alpes, IPAG, F-38000 Grenoble, France\label{grenoble1}
\and CNRS, IPAG, F-38000 Grenoble, France\label{grenoble2}
\and Instituto de Astrof\'isica e Ci\^{e}ncias do Espa\c co, Universidade do Porto, CAUP, Rua das Estrelas, PT4150-762 Porto, Portugal\label{porto2}
\and Stellar Astrophysics Centre, Department of Physics and Astronomy, Aarhus University, Ny Munkegade 120, DK-8000 Aarhus C, Denmark\label{aarhus}
\and Aix Marseille Universit\'e, CNRS, LAM (Laboratoire d'Astrophysique de Marseille) UMR 7326, 13388, Marseille, France\label{lam}
\and Observatoire Astronomique de l'Universit\'e de Gen\`eve, 51 chemin des Maillettes, 1290 Versoix, Switzerland\label{geneva}
\and Observatoire de Haute Provence, 04670 Saint Michel l'Observatoire, France\label{ohp}
\and Institut d'Astrophysique de Paris, UMR7095 CNRS, Universit\'e Pierre \& Marie Curie, 98bis boulevard Arago, 75014 Paris, France\label{iap}
\and Departamento de F\'{i}sica, Universidade Federal do Rio Grande do Norte, 59072-970 Natal, RN, Brazil\label{natal}
\and Departamento de Astronom\'{i}a, Universidad de Chile, Santiago, Chile\label{santiago}
\and Departamento de F\'{i}sica e Astronomia, Faculdade de Ci\^encias, Universidade do Porto, Rua do Campo Alegre 687, PT4169-007 Porto, Portugal\label{porto3}
}

   \date{}

   \date{}

% \abstract{}{}{}{}{} 
% 5 {} token are mandatory
 
  \abstract
  {\ktwo\ space observations recently found that three super-Earths transit 
the nearby M dwarf \epic. The apparent brightness and the small physical 
radius of their host star rank these planets amongst the most favourable 
for follow-up characterisations. The outer planet orbits close to the 
inner edge of the habitable zone and might become one of the first 
exoplanets searched for biomarkers using transmission spectroscopy. 
We used the HARPS velocimeter to measure the mass of the planets. The mass of 
planet~$b$ is $8.4\pm2.1$~M$_\oplus$, while our determination of those planets~$c$ and $d$ are affected by the stellar activity. With a density of 
$4.32^{+2.0}_{-0.76}$~$\mathrm{g\;cm^{-3}}$, planet~$b$ is probably 
mostly rocky, but it could contain up to 50\% water.}

   \maketitle
%
%________________________________________________________________

\section{Introduction}

\citet{2015arXiv150103798C} have recently reported from \ktwo\ 
\citep{2014PASP..126..398H} observations that the star \epic\ 
(2MASS 11292037-0127173, EPIC~201367065) hosts a planetary system 
with three transiting super-Earths. The star is an inactive M0 dwarf 
($T_{\mathrm{eff}} = 3900 \pm 190$ K, [Fe/H]$= -0.32 \pm 0.13$ dex), 
and it is bright enough to be amenable to transit spectroscopy. 
With a radius of 1.5~R$_\oplus$, Planet~$d$ orbits close to the inner edge 
of the system's habitable zone. The mass of the planets has not been 
measured to date. 

To constrain those masses, we monitored the radial velocity of \epic\
with the HARPS velocimeter \citep{2003Msngr.114...20M}. We jointly
analysed these new velocities and the \ktwo\ photometry through n-body 
integrations to obtain the mass of planets. Combined with the planet 
sizes, this constrains their mean densities and bulk compositions and helps 
explore the rocky--gaseous transition in the super-Earth regime. 

\section{Observations}
\subsection{\ktwo\ light curve}

The \ktwo\ mission observed \epic\ during its Campaign 1 (Summer 2014) 
in long cadence mode\footnote{http://archive.stsci.edu/k2/}.  We 
downloaded the pixel data from the  Mikulski Archive for Space Telescopes 
(MAST)\footnote{http://archive.stsci.edu/kepler/data\_search/search.php} 
and used a modified version of the CoRoT imagette pipeline 
\citep{2014A&A...569A..74B} to extract a light curve. The procedure 
first determines the circular synthetic aperture 
that maximises the photometric signal-to-noise ratio on the mean image. For each image, it then computes 
a modified moment method centroid \citep{1989AJ.....97.1227S} 
and recentres a heavily oversampled version of the image on the centroid
before extracting the flux inside the pre-determined aperture. The degraded 
pointing stability of the \ktwo\ mission couples with pixel sensitivity 
variations to introduce position-dependent systematics in the raw light 
curves. To correct for this flux dependence with position, we used a 
procedure inspired by \citet{2014PASP..126..948V}. The light curve was 
divided into seven equal duration segments, and for each segment we 
performed a 1D decorrelation as described in \citet{2014arXiv1412.1827V}. 
We found that a 21-pixel 
photometric aperture results in the best photometric precision of the 
final light curve with a 204~ppm mean RMS. The 80-day light curve 
contains eight transits of Planet~$b$, four of Planet~$c$, and two of Planet~$d$. 
We only modelled the light curve around those transits, after normalising it 
with a parabolic fit to its out-of-transit part. To account for the 
29.4~minutes integration time of the long cadence data, we oversampled 
the model light curve by a factor of 20 and then binned it to the cadence 
of the data points.

\subsection{HARPS radial velocities}
\label{harpsrv}

We obtained differential radial velocities (RVs) with HARPS, the ESO 
velocimeter installed at the focus of the 3.6m telescope at La Silla 
Observatory (Mayor et al. 2003; Pepe et al. 2003). We chose not to use 
simultaneous wavelength calibration and to instead rely on the $<$1~\ms\ 
stability of the HARPS spectrograph, since we expected significantly larger
photon noise errors. We tried to secure two 1800~s observations per night  
and collected 66~spectra over a timespan of 103~days. 

For an optimal extraction of the velocity signal, we used all recorded 
spectra \citep[already extracted and wavelength-calibrated by the 
online pipeline,][]{2007A&A...468.1115L}
to build stellar templates. We shifted each spectrum by its barycentric
correction and co-added those shifted spectra into higher S/N templates after 
carefully masking the telluric absorption lines. 

When using such a small set of relatively noisy spectra, we would bias the velocity if we included the spectrum analysed for Doppler shift in the template that it is matched against, since their common noise pattern will contribute to the match. To avoid this bias, we built one template for each spectrum 
by co-adding all spectra {\it \emph{but}} the spectrum itself.

We then computed the RV shifts that minimise the $\chi^2$ difference between 
the individual spectra and their templates 
\citep[e.g.][]{2006MNRAS.371.1513Z,2012ApJS..200...15A,2014arXiv1411.7048A}. 
The 66 resulting velocities, listed in Table~\ref{table_rv}, have 
a 4.9~\ms\ dispersion for a median uncertainty of 2.9~\ms\ (compared 
with 7.1~\ms\ and 4.3~\ms\ for the velocities measured
by the HARPS pipeline). Although the orbital periods are known from 
the photometry and do not need to be identified from the RVs alone, 
Fig.~\ref{periodogram} shows their periodogram for illustration.

\subsection{Stellar activity}
\label{stellaractivity}

The HARPS spectral range includes both H$\alpha$ and the 
\ion{Ca}{ii}~H\&K lines, which are good tracers of stellar activity. 
While H$\alpha$ is a pure absorption line in \epic\ and does not measurably 
vary, its \ion{Ca}{ii} lines, as for all M dwarfs, do have emission 
cores. We quantify the \ion{Ca}{ii} ~H\&K  flux through the S index 
\citep{1978PASP...90..267V} and find that it varies by 30\% on a
timescale that is longer than than the ten-day period of Planet~$b$. 
\epic\ has a lower average S index than Gl 846, an M0.5V star with a 
$\sim$10.6-day rotation period \citep{2013A&A...549A.109B}, 
consistently with $P_{\rm rot}>10$ days. We used our recent calibration 
(Astudillo-Defru et al. in prep.) to compute the $R^\prime_{HK}$ index 
from S, and from  the $R^\prime_{HK}$ vs $P_{\rm rot}$ in the same paper, we
estimate $P_{\rm rot}\simeq 38$~days. The \ktwo\ lightcurve varies with 
a \textless~2~mmag semi-amplitude and a $\sim$40-day characteristic 
timescale, consistent with a fairly inactive star and with the estimated
$P_{\rm rot}$. If that photometric variability is entirely due to a 
dark spot rotating in and out of view, the corresponding RV semi-amplitude 
is at most $\sim$1.6~\ms. 

In an effort to quantify activity-induced RV variations, we developed 
new activity metrics. We built an "active" and a "non-active" template 
from the third of the observed spectra with the highest and 
the lowest S-indices, respectively, computed two sets of active and non-active RVs,
and used their difference $\Delta$RV=RV$_{\rm act}-$RV$_{\rm noact}$ as an activity
tracer. The $\Delta$RV time series (Fig.~\ref{figRV}) varies by $\pm8$ m/s 
for BJD$ - 2~450~000 < 7070$, and is consistent with zero after that 
date. We surmise that the early large variations result from a spot 
that not only occults part of the star but also imprints the overall 
spectrum with its spectral signature. That spot would have been present 
during the first $\sim25$ days of the observations and then disappeared.

We computed our nominal RVs (Sect.~\ref{harpsrv}) with a template that
includes all spectra regardless of their S index, and it presumably
has a sensitivity to activity that is intermediate between that of the 
active and non-active template. We account for its stellar activity 
sensitivity by introducing a term proportional to $\Delta RV$ in the
radial velocity model, with the proportionality coefficient $\alpha$ adjusted
as a free parameter in the fit.

\section{Analysis}

We analyse this multi-planetary system with an n-body dynamical model 
that describes the gravitational interactions between all components of 
the system and not just the pull of the central star. Unlike a 
Keplerian model, the dynamical model therefore accounts for the photometric 
and radial-velocity changes induced by the  mutual attraction of the 
planets, such as transit timing variations (TTVs), transit duration 
variations (TDVs), or more generally, transit shape variations. 
Even though we detect no gravitational interactions, using a dynamical 
model helps constrain the masses and eccentricities by excluding  
values for those parameters that would result in detectable 
interactions or in a highly unstable system.

Our dynamical model uses the well known 
\textsc{mercury} \citep{1999MNRAS.304..793C} n-body integrator code to 
compute the three-dimensional position and velocity of all system components 
as a function of time. The line-of-sight projection of the stellar velocity 
is compared to the HARPS radial velocities. The sky-projection of the 
planet-star separations are used, together with the planet-to-star-radius 
ratios and the limb darkening coefficient, to model the \ktwo\ light curve 
\citep{2002ApJ...580L.171M}. We couple the dynamical model with a Monte 
Carlo Markov Chain (MCMC) code, described in detail in 
\citet{2014MNRAS.441..983D}, to explore the posterior distribution of the 
parameters. For each MCMC step, we run \textsc{mercury} three times with 
the same model parameters: 
\begin{list}{-}{}
\item{} once over the 80~days of the \ktwo\ observations, using the 
Bulirsch-Stoer algorithm and a 0.02-day timestep to model the 
light curve with a 1~ppm maximum photometric error
\item{} once over the (disjoint) 103~days of the HARPS observations,
again with the Bulirsch-Stoer algorithm and a 0.02~days timestep, to model 
the radial velocities
\item{} once over a 1000-year interval, using the hybrid 
symplectic/Bulirsch-Stoer integrator with a 0.5-day timestep 
(1/20th of the shortest orbital period)
\end{list}
The last run is not compared to any observation but used to ensure the 
short-term stability of the system, while rejecting any MCMC step where two 
orbits intersect or where a planet comes within 0.05 au of the star. (Tidal forces are not included in \textsc{mercury}.) One would ideally 
like to evaluate stability on longer time scales, with 1000~years 
chosen as a compromise between ideality and computational expense.

The physical parameters of the model are the stellar mass and radius, the 
coefficient of a linear limb-darkening law, the systemic velocity, the 
planetary masses, the planet-to-star radius ratios, and the planetary 
orbital parameters ($a$, $e$, $i$, $\omega$, $n$, and $M$; see 
Table~\ref{table}) at reference time 2~456~812~BJD. To minimise 
correlations, however, we use a different parametrisation for MCMC 
jump parameters: orbital period ($P$), conjunction time of the first 
transit observed by \ktwo\ ($T_0$), radial velocity semi-amplitude 
($K$), $\sqrt{e}\cos{\omega}$, and $\sqrt{e}\sin{\omega}$. Additionally, we fit a global light curve normalization factor, one multiplicative jitter parameter for each data set and  
$\alpha$, the amplitude of the $\Delta$RV term of the radial
velocity model (Sec.~\ref{stellaractivity}).

We use non-informative uniform priors for all MCMC model parameters except 
the stellar mass and radius, for which we adopt the Gaussian distributions 
of \citet{2015arXiv150103798C} (we also adopt all parameters in common as 
the starting point of our chains) and the longitude of the ascending node 
for which we adopt a Gaussian distributions with $\sigma = 2\degree$ to
enforce the observed physical bias towards coplanarity 
\citep{2014ApJ...790..146F}. We adopt a non-informative 
prior for the limb-darkening even though this widens the distributions 
of several parameters, because limb-darkening models have often been 
found to be inaccurate when the data quality is high enough to probe them. 
We ran 48 MCMC chains of 180~000 steps and combined their results as 
described in \citet{2014MNRAS.441..983D}.

\section{Results}
\label{results}

\begin{figure}[!ht]
\includegraphics[width=9cm]{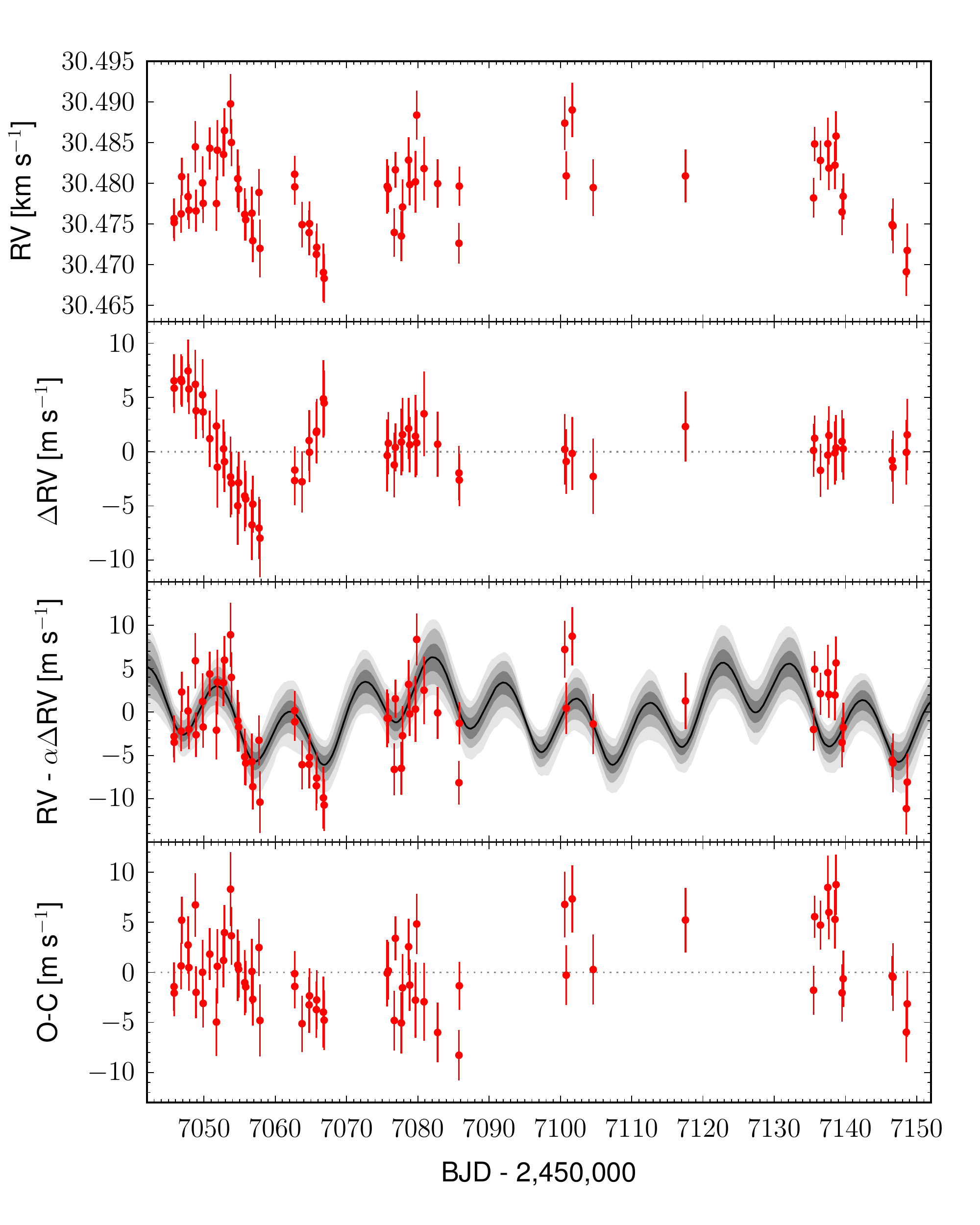}
\caption{Time series for, from top to bottom: a) the HARPS radial velocities 
of \epic, b) $\Delta$RV (Sec.~\ref{stellaractivity}), c) the RVs corrected 
from the $\alpha\Delta$RV of the maximum likelihood model together with 
the dynamical model (the solid black line represents the median model, 
and the shades of grey represent the 68.3, 95.5, and 99.7\% Bayesian confidence intervals), and d) the RV residual from the maximum likelihood dynamical model.}
\label{figRV}
\end{figure}

Table~\ref{table} lists the mode and the 68.3\% confidence interval for 
the system parameters. Figures~\ref{figRV} and \ref{figPH} show the radial 
velocity measurements and the transit light curves, together with their 
respective models. By fitting a stellar spectral energy distribution 
to the \epic\ photometry \citep[][Table~1]{2015arXiv150103798C} as 
described in \citet{2014MNRAS.441..983D}, we obtain a 47.5~$\pm$~6.0 pc
distance and $E_{(B-V)}$ < 0.056 with 68.3\% confidence. 

Our results agree with \citet{2015arXiv150103798C} for all parameters
in common. In addition to adding the radial velocity information, our 
analysis differs in that we neither impose a circular orbit nor use 
informative priors on $R_\star/a$ or the limb darkening coefficient.
We also model the transits of all three planets jointly rather than 
one planet at a time, enforcing consistency in the derived stellar
properties. Our approach measures the stellar density more accurately 
than can be inferred from the average spectrum.

The times of the individual transit are measured with standard errors 
(which roughly translate to a 1$\sigma$ detection limits for TTVs) of 
1.4, 2.8, and 2.9~minutes for planets~$b$, $c$, and $d$, respectively. 
Over the short time span of the HARPS observations, the radial velocity 
of the dynamical model differs from that of a three-Keplerian model by
at most $27\cms$, which is well below the ${\sim}2\ms$ measurement noise.

Owing to the low eccentricities and almost edge-on inclinations, all 
three planets almost certainly undergo secondary eclipses. Table~\ref{table}
lists the epochs and durations of these secondary eclipses.

\begin{figure}[!ht]
\includegraphics[width=8.5cm]{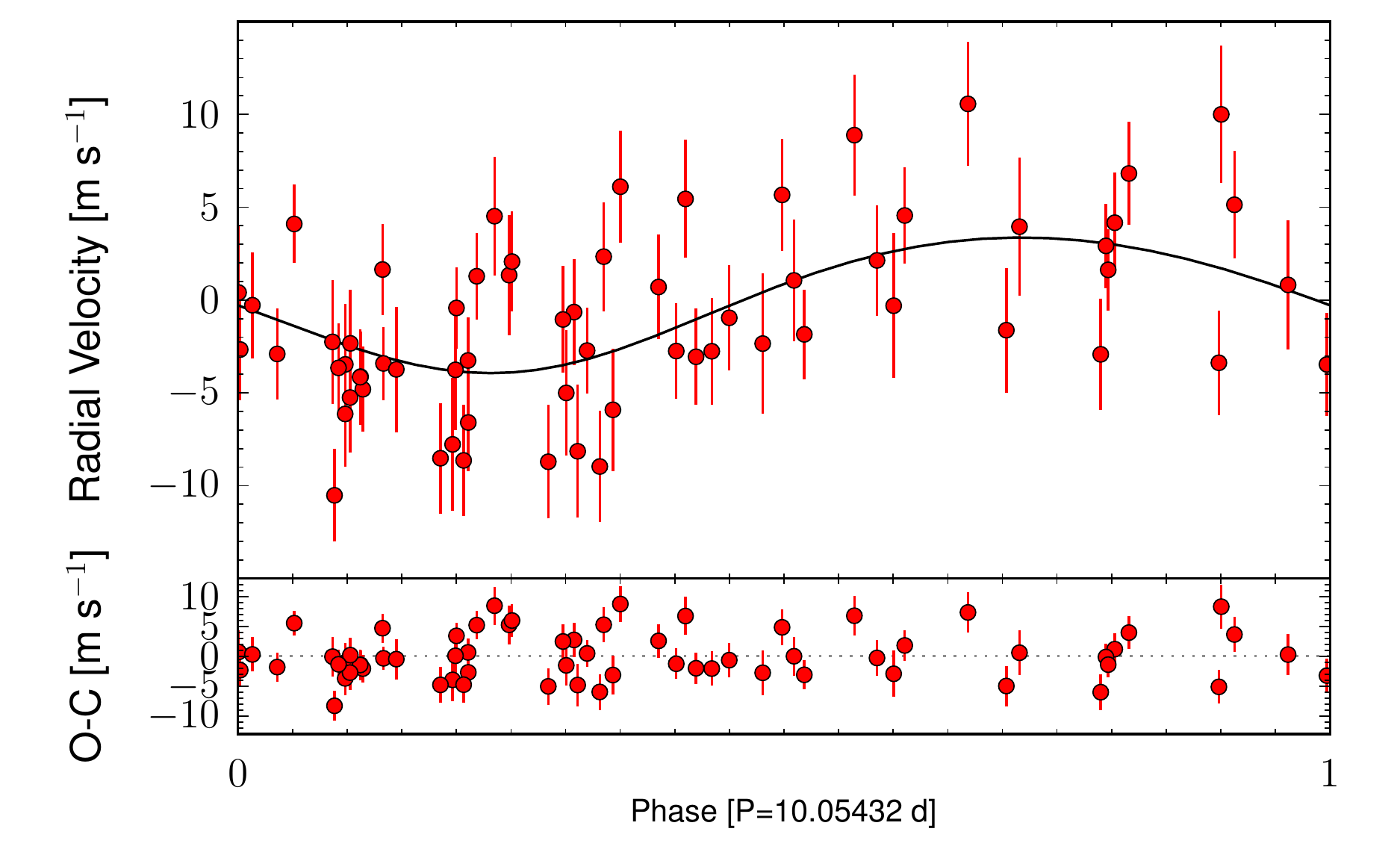}
\caption{HARPS radial velocities of \epic\ phased to the period of Planet~$b$,
with the Keplerian model (solid black line) overlaid, after removal of the 
Keplerians signals from Planets~$c$ and $b$ and of $\alpha\Delta$RV activity
term of the maximum likelihood model.}
\label{RVphase}
\end{figure}

To evaluate the robustness of our mass measurements, we extended the radial 
velocity by adding a linear drift, a fourth planet, a term proportional
to the bisector of the spectra, and any combination of them. We also 
experimented with restricting the analysis to those radial velocities 
that seem unaffected by stellar activity (BJD$-2~450~000>7070$). They
always produced similar velocity amplitudes (and masses) for Planet~$b$,
but a wide range of values for Planets~$c$ and $d$.  We conclude that 
our data robustly measure the mass of Planet~$b$, $8.4\pm2.1$~\Mearth, 
but not those of $c$ and $d$. Many more observations and very careful
analysis will be needed to disentangle signatures of Planets~$c$ and $d$ 
from the activity signal. 
Figure~\ref{RVphase} shows the RV signal from Planet~$b$ after removing the other terms of the model.

When included in the model, the fourth planet converges to a $P_e\sim100$-day
period and absorbs part of the residuals seen in Fig.~\ref{figRV} (bottom).
Model comparison favours the four-planet model only marginally over the 
simpler three-planet model, with the Bayesian evidence estimators of
\citet{perrakis2014} and \citet{chibjeliazkov2001} giving odd ratios 
of 4.6$\pm$0.3 and 4.9$\pm$2.1, respectively, in favour of the four-planet 
model. More data will thus be needed to firmly establish whether
additional planets orbit \epic.

\begin{figure}
  \centering
\includegraphics[width=8cm]{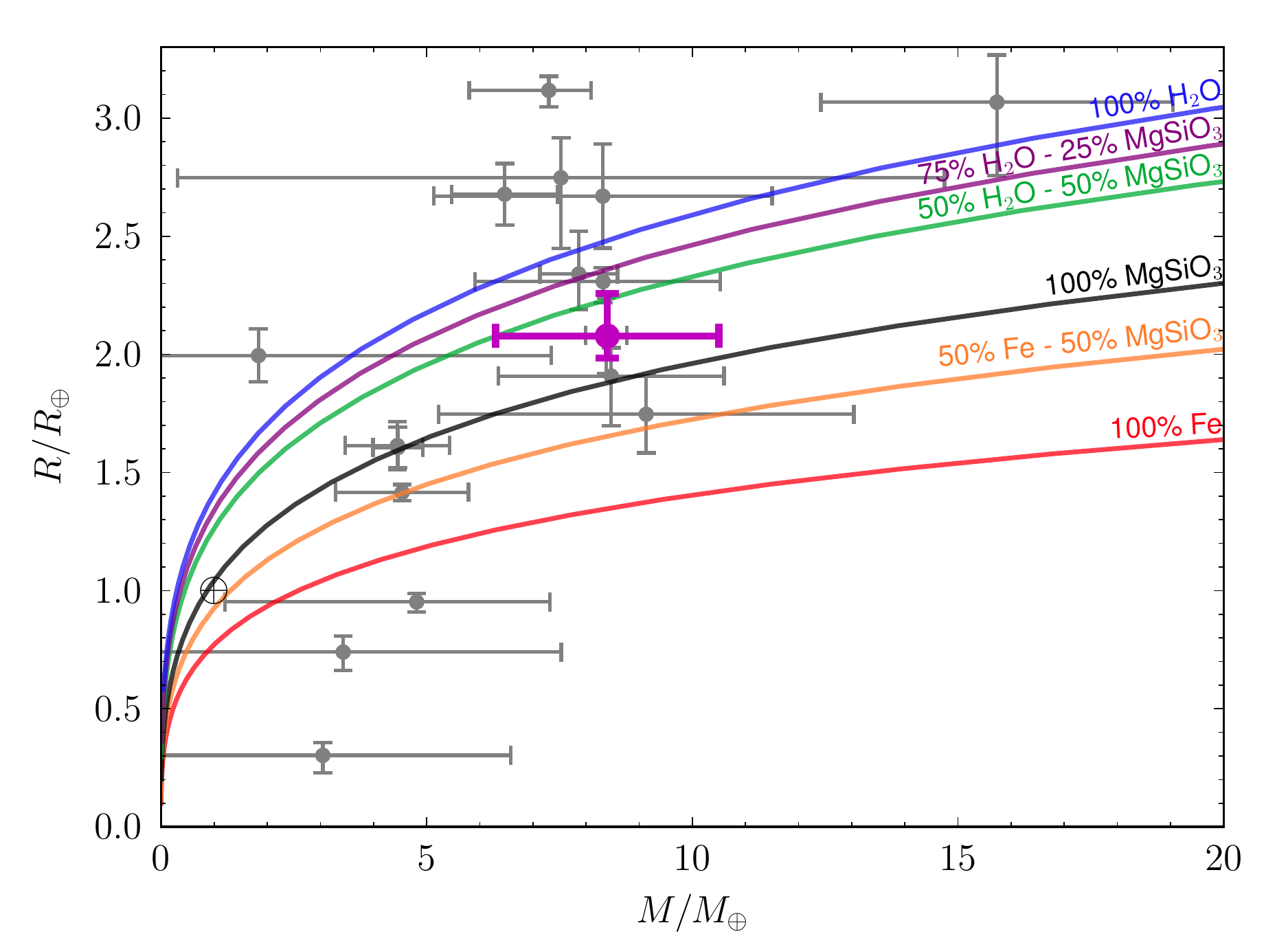}
\caption{Radius versus mass diagram of the known small exoplanets 
\citep{2011PASP..123..412W}, with planet~$b$ added (magenta) with 
68.3\% credible intervals. The colored solid lines represents 
theoretical models for differents compositions
\citep{2013PASP..125..227Z}.}
\label{figMR}
\end{figure}

Figure~\ref{figMR} adds Planet~$b$ to the mass-radius 
diagram of the known small transiting planets. Planet~$b$ must contain
rock with at most a 50\% water envelope.

Planets around M stars are ideal for characterisation follow-up, thanks to 
the favourable planet-to-star surface ratio. The other known low-mass planets 
transiting bright M stars have a gas envelope 
\citep[GJ436b and GJ3470b,][]{2004ApJ...617..580B,2012A&A...546A..27B} 
or may even potentially consist of 100\% water 
\citep[GJ1214b,][]{2009Natur.462..891C}. Our measurement of the density of 
Planet~$b$ shows that it is either entirely rocky or mostly rocky with a water 
envelope. It is the first planet in this category, and the brightness of the star makes it a prime target for follow-up.

\begin{acknowledgements}
We thank the ESO La Silla staff for its continuous support, the HARPS 
observers (C. Mordasini, J. Martins, and A. Sozzetti, as well as 
L. Kreidberg for her Mandel \& Agol code), 
C. Damiani for discussions of the dynamics of the system, and 
T. Fenouillet for his assistance with the LAM computing cluster.
 
This paper includes data collected by the \Kepler\ mission. Funding 
for the \Kepler\ mission is provided by the NASA Science Mission 
directorate. Some of the data presented in this paper were obtained 
from the Mikulski Archive for Space Telescopes (MAST). STScI is 
operated by the Association of Universities for Research in Astronomy, 
Inc. under NASA contract NAS5-26555. Support for MAST for non-HST data 
is provided by the NASA Office of Space Science via grant NNX09AF08G 
and by other grants and contracts.
This research made use of the Exoplanet Orbit Database and the Exoplanet 
Data Explorer at exoplanets.org.
N. A.-D. acknowledges support from CONICYT Becas-Chile number 72120460. 
XB and JMA acknowledge funding from the European Research Council under 
the ERC Grant Agreement n. 337591-ExTrA. 
NCS acknowledges support by  Funda\c{c}\~ao para a Ci\^encia e a 
Tecnologia (FCT) through Investigador FCT contracts of reference 
IF/00169/2012, and POPH/FSE (EC) by FEDER funding through the program 
``Programa Operacional de Factores de Competitividade - COMPETE''. 
A.S. is supported by the European Union under a Marie Curie Intra-European 
Fellowship for Career Development with reference FP7-PEOPLE-2013-IEF, number 
627202. SCCB and OD thank the CNES for their grants 98761 and 124378, respectively
.
\end{acknowledgements}

\bibliographystyle{aa}
\bibliography{K2-3}

%-------------------------------------------------------------------

\Online
\begin{appendix}
\section{Additional figures and tables}

\begin{table}
\small
  \caption{HARPS radial velocity measurements of \epic.}
\begin{tabular}{lccccc}
\hline
\hline
BJD - 2\,400\,000 & RV & $\pm1\sigma^{\dagger}$ & $\Delta$RV & S-index\\
& [km\,s$^{-1}$] & [km\,s$^{-1}$]  & [m\,s$^{-1}$] &    \\
\hline
57045.795776    & 30.4757       & 0.0024 &  6.5 & 0.769          \\
57045.816981    & 30.4752       & 0.0023 &  5.9 & 0.724  \\
57046.785448    & 30.4762       & 0.0023 &  6.7 & 0.775  \\
57046.865234    & 30.4808       & 0.0023 &  6.5 & 0.782  \\
57047.761529    & 30.4784       & 0.0029 &  7.5 & 0.742  \\
57047.881539    & 30.4767       & 0.0023 &  5.8 & 0.731  \\
57048.787741    & 30.4845       & 0.0032 &  6.2 & 0.696  \\
57048.884890    & 30.4766       & 0.0026 &  3.8 & 0.826  \\
57049.786450    & 30.4800       & 0.0033 &  5.3 & 0.629  \\
57049.881769    & 30.4775       & 0.0024 &  3.7 & 0.777  \\
57050.806327    & 30.4843       & 0.0026 &  1.2 & 0.653  \\
57051.741719    & 30.4775       & 0.0034 &  2.4 & 0.520  \\
57051.862734    & 30.4841       & 0.0037 & -1.4 & 0.779  \\
57052.740065    & 30.4835       & 0.0027 &  0.3 & 0.659  \\
57052.870155    & 30.4865       & 0.0028 & -0.9 & 0.783  \\
57053.720306    & 30.4898       & 0.0037 & -2.3 & 0.657  \\
57053.841750    & 30.4850       & 0.0029 & -2.9 & 0.670  \\
57054.727180    & 30.4806       & 0.0036 & -5.0 & 0.779  \\
57054.854087    & 30.4793       & 0.0029 & -2.9 & 0.666  \\
57055.709838    & 30.4762       & 0.0032 & -4.1 & 0.660  \\
57055.846433    & 30.4755       & 0.0026 & -4.4 & 0.720  \\
57056.724361    & 30.4763       & 0.0032 & -6.8 & 0.570  \\
57056.842204    & 30.4729       & 0.0026 & -4.8 & 0.739  \\
57057.713776    & 30.4789       & 0.0029 & -7.0 & 0.757  \\
57057.849294    & 30.4720       & 0.0036 & -8.0 & 0.822  \\
57062.713162    & 30.4811       & 0.0023 & -2.7 & 0.889  \\
57062.733973    & 30.4796       & 0.0022 & -1.7 & 0.890  \\
57063.753092    & 30.4749       & 0.0028 & -2.8 & 0.838  \\
57064.748274    & 30.4739       & 0.0028 &  1.0 & 0.976  \\
57064.796239    & 30.4750       & 0.0027 & -0.1 & 0.865  \\
57065.762656    & 30.4713       & 0.0028 &  1.8 & 0.874  \\
57065.807820    & 30.4721       & 0.0030 &  1.9 & 0.896  \\
57066.751931    & 30.4690       & 0.0036 &  4.9 & 0.781  \\
57066.853522    & 30.4683       & 0.0030 &  4.5 & 0.943  \\
57075.700364    & 30.4796       & 0.0033 & -0.4 & 0.992  \\
57075.864582    & 30.4793       & 0.0029 &  0.8 & 1.007  \\
57076.694658    & 30.4740       & 0.0030 & -1.2 & 0.015  \\
57076.841851    & 30.4817       & 0.0022 &  0.4 & 0.980  \\
57077.687319    & 30.4735       & 0.0031 &  0.9 & 1.023  \\
57077.852660    & 30.4771       & 0.0034 &  1.6 & 0.015  \\
57078.702537    & 30.4829       & 0.0028 &  2.1 & 0.950  \\
57078.865482    & 30.4798       & 0.0025 &  0.6 & 0.918  \\
57079.661385    & 30.4802       & 0.0038 &  1.4 & 0.015  \\
57079.839990    & 30.4884       & 0.0030 &  0.8 & 0.014  \\
57080.867369    & 30.4818       & 0.0039 &  3.5 & 0.015  \\
57082.773669    & 30.4800       & 0.0030 &  0.7 & 0.939  \\
57085.773295    & 30.4726       & 0.0025 & -2.0 & 0.780  \\
57085.810298    & 30.4797       & 0.0024 & -2.6 & 0.786  \\
57100.614169    & 30.4874       & 0.0033 &  0.2 & 0.774  \\
57100.822420    & 30.4809       & 0.0030 & -0.9 & 0.822  \\
57101.660042    & 30.4890       & 0.0034 & -0.2 & 0.576  \\
57104.607546    & 30.4795       & 0.0035 & -2.3 & 0.932  \\
57117.543836    & 30.4809       & 0.0032 &  2.3 & 0.545  \\
57135.517340    & 30.4782       & 0.0024 &  0.1 & 0.838  \\
57135.673963    & 30.4848       & 0.0021 &  1.2 & 0.959  \\
57136.486899    & 30.4828       & 0.0024 & -1.7 & 0.742  \\
57137.518119    & 30.4849       & 0.0032 & -0.3 & 0.866  \\
57137.679348    & 30.4819       & 0.0027 &  1.5 & 0.659  \\
57138.523321    & 30.4822       & 0.0029 & -0.1 & 0.867  \\
57138.676714    & 30.4858       & 0.0030 &  0.4 & 0.839  \\
57139.518338    & 30.4765       & 0.0029 &  1.0 & 0.835  \\
57139.679473    & 30.4784       & 0.0028 &  0.3 & 0.813  \\
57146.549907    & 30.4749       & 0.0020 & -0.8 & 0.825  \\
57146.668220    & 30.4745       & 0.0034 & -1.4 & 0.687  \\
57148.541552    & 30.4691       & 0.0030 & -0.1 & 0.946  \\
57148.662503    & 30.4717       & 0.0033 &  1.6 & 0.619  \\
\hline
\end{tabular}
\begin{list}{}{}
\item {\bf{Notes.}} $^{(\dagger)}$ include a systematic error of 60~\cms\ \\\citep{2013A&A...549A.109B}.
\end{list}
\label{table_rv}
\end{table}

\begin{figure}[!ht]
\centering
\vspace{-0.3cm}
\includegraphics[width=8cm]{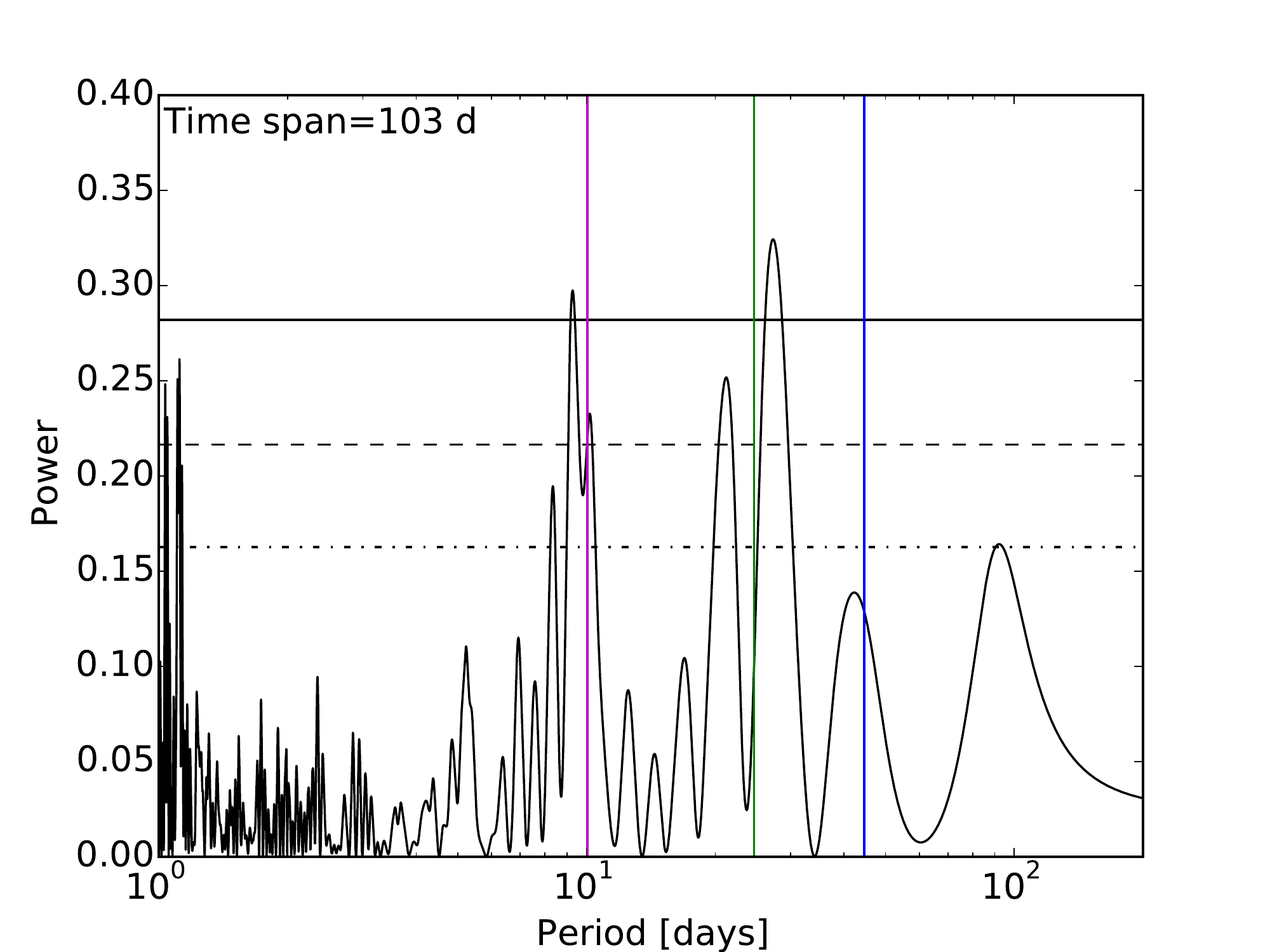}
\caption{Periodogram of HARPS radial velocities of \epic. The horizontal lines correspond to 1, 2, and 3$\sigma$ confidence intervals. The false alarm probabilities (FAP) of the main peaks are 0.044\% (20-30~days) and 0.0015\% (8-10~days). The vertical lines indicate the period of Planets~$b$, $c,$ and $d$ (from left to right).}
\label{periodogram}
\end{figure}

%---------------------

\begin{table*}
\tiny
  \renewcommand{\arraystretch}{0.7}
\centering
\cprotect\caption{Model parameters. Posterior mode and 68.3\% credible 
intervals. The orbital elements have their origin at the star 
({\it Asteroidal parameters} in the \textsc{mercury} code) and are 
osculating elements for reference time $t_{\mathrm{ref}} = 2~456~812$ BJD. 
We have low 
confidence in the mass and density of planets~$c$ and $d$ 
(Sec.~\ref{results}).}\label{table}
\begin{tabular}{lccc}
\hline \\ [-1ex]
Parameter & \multicolumn{2}{c}{Mode and 68.3\% credible interval} \smallskip\\
\hline \\ [-1ex]
Stellar mass, $M_\star$ [\Msun]$^{\bullet}$         & 0.612 $\pm$ 0.086 & & \\
Stellar radius, $R_\star$ [\Rsun]$^{\bullet}$       & 0.553 $\pm$ 0.041 & & \\
Stellar density, $\rho_{\star}$ [$\rho_\odot$]      & 3.58 $\pm$ 0.61 & & \\
Surface gravity, \logg\ [cgs]                     & 4.734 $\pm$ 0.062 &  & \\
Linear limb darkening coefficient, $u^{\bullet}$ & 0.573 $\pm$ 0.088 &  & \\
Systemic velocity, $\gamma_0$ [\kms]$^{\bullet}$   & 30.48024 $\pm$ 0.00063 &  & \medskip\\

\multicolumn{1}{l}{} & \emph{Planet~$b$} & \emph{Planet~$c$} & \emph{Planet~$d$} \smallskip\\
Semi-major axis, $a$ [AU]                                 & 0.0775 $\pm$ 0.0039 & 0.1405 $\pm$ 0.0067   & 0.2086 $\pm$ 0.010 \\
Eccentricity, $e$    & [0, 0.12]$^{\ast}$ < 0.29$^{\ddagger}$ & [0, 0.08]$^{\ast}$ < 0.20$^{\ddagger}$  & [0, 0.09]$^{\ast}$ < 0.19$^{\ddagger}$ \\
Inclination, $i$ [\degree]$^{\bullet,\dagger}$              & 89.59$^{+0.24}_{-0.40}$  & 89.70 $\pm$ 0.20    & 89.79 $\pm$ 0.15 \\
Argument of pericentre, $\omega$ [\degree]                & 180 $\pm$ 70          & 89 $\pm$ 110        & 351 $\pm$ 66 \\
Longitude of the ascending node, $n$ [\degree]$^{\bullet}$ & 180 (fixed)           & 180.0 $\pm$ 1.8     & 180.0 $\pm$ 1.8 \\
Mean anomaly, $M$ [\degree]                               & 230 $\pm$ 69          & 21 $\pm$ 100        & 338 $\pm$ 67 \smallskip\\

Period, $P$ [days]$^{\bullet}$                        & 10.05429 $\pm$ 0.00022      & 24.6491 $\pm$ 0.0033   & 44.5705 $\pm$ 0.0059 \\
Transit epoch, $T_0$ [BJD-2,450,000]$^{\bullet}$      & 6813.41817 $\pm$ 0.00082    & 6812.2784 $\pm$ 0.0018 & 6826.2272 $\pm$ 0.0027 \\
Transit duration, $T_{14}$ [hours]                   &  2.563 $\pm$ 0.039          &  3.345 $\pm$ 0.079  & 4.07 $\pm$ 0.12 \\
Scaled semi-major axis, $a/R_{\star}$                & 30.74$^{+0.53}_{-2.5}$         & 55.90$^{+0.97}_{-4.6}$  & 82.9$^{+1.4}_{-6.9}$ \\
Transit impact parameter, $b$                       & 0.21$^{+0.21}_{-0.12}$         & 0.31$^{+0.14}_{-0.21}$   & 0.359$^{+0.089}_{-0.27}$ \\
Secondary eclipse impact parameter, $b_{SE}$         & 0.310$^{+0.079}_{-0.21}$       & 0.33$^{+0.11}_{-0.22}$   & 0.30$^{+0.14}_{-0.20}$ \\
Secondary eclipse epoch, $T_{SE}$ [BJD-2,450,000]    & 6818.407$^{+0.066}_{-0.79}$   & 6824.59 $\pm$ 0.85       & 6848.72$^{+2.9}_{-0.24}$ \\
Secondary eclipse duration [hours]                  & 2.55 $\pm$ 0.25              & 3.37$^{+0.40}_{-0.17}$     & 4.01$^{+0.41}_{-0.21}$ \\
Radial velocity semi-amplitude, $K$ [\ms]$^{\bullet}$ & 3.60 $\pm$ 0.87              & 0.61$^{+0.66}_{-0.37}$    & 2.84 $\pm$ 0.90  \\
Radius ratio, $R_p/R_\star^{\bullet}$                & 0.03477 $\pm$ 0.00064        & 0.02770 $\pm$ 0.00070   & 0.02495 $\pm$ 0.00074 \\
$\sqrt{e}\cos{\omega}^{\bullet}$                    & -0.283$^{+0.33}_{-0.065}$      & 0.01 $\pm$ 0.19         & 0.263$^{+0.052}_{-0.28}$  \\
$\sqrt{e}\sin{\omega}^{\bullet}$                    & -0.05$^{+0.21}_{-0.14}$        & 0.128$^{+0.10}_{-0.26}$   & 0.00 $\pm$ 0.16 \smallskip\\

Planet mass, $M_{p}$ [\Mearth]                   & 8.4 $\pm$ 2.1          & 2.1$^{+2.1}_{-1.3}$    & 11.1 $\pm$ 3.5 \\
Planet radius, $R_{p}$[\Rearth]                  & 2.078$^{+0.18}_{-0.093}$ & 1.644$^{+0.16}_{-0.065}$ & 1.53 $\pm$ 0.11 \\
Planet density, $\rho_p$ [$\mathrm{g\;cm^{-3}}$] & 4.32$^{+2.0}_{-0.76}$    & 1.82$^{+3.0}_{-0.96}$  & 17.5 $\pm$ 6.3 \medskip\\

\ktwo\ long-cadence multiplicative jitter$^{\bullet}$ & 1.020 $\pm$ 0.049 & & \\
HARPS multiplicative jitter$^{\bullet}$               & 1.43 $\pm$ 0.13   & & \\
Amplitude of the $\Delta$RV term, $\alpha^{\bullet}$             & -0.21 $\pm$ 0.19   & & \smallskip\\

\hline
\end{tabular}
\begin{list}{}{}
\item {\bf{Notes.}} $^{(\bullet)}$ MCMC jump parameter. $^{(\dagger)}$ reflected 
with respect to $i = 90\degree$, the supplementary angle is equally 
probable. 
$^{(\ast)}$ 68.3\% highest density interval. 
$^{(\ddagger)}$ upper limit, 99\% confidence. 
$T_0 \equiv t_{\mathrm{ref}} - \frac{P}{2\pi}\left(M-E+e\sin{E}\right)$ 
with $E=2\arctan{\left\{\sqrt{\frac{1-e}{1+e}}\tan{\left[\frac{1}{2}\left(\frac{\pi}{2}-\omega\right)\right]}\right\}}$, $P \equiv \sqrt{\frac{4\pi^2a^{3}}{G M_{\star}}}$, $K \equiv \frac{M_p \sin{i}}{M_\star^{2/3}\sqrt{1-e^2}}\left(\frac{2 \pi G}{P}\right)^{1/3}$. 
$\Msun$~=~1.98842\ten[30]~kg, \Rsun~=~6.95508\ten[8]~m, 
\Mearth~=~5.9736\ten[24]~kg, \Rearth~=~6,378,137~m.
\end{list}
\end{table*}

\begin{figure*}[!ht]
\includegraphics[height=3.1cm]{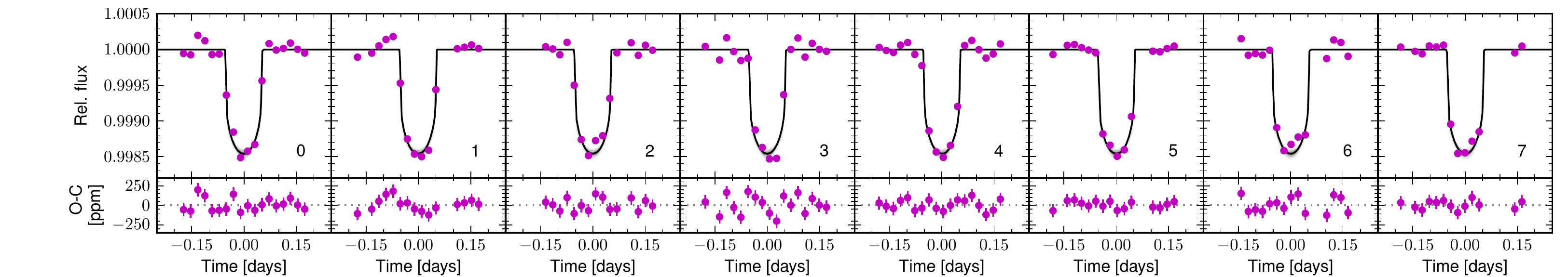}\\
\includegraphics[height=3.1cm]{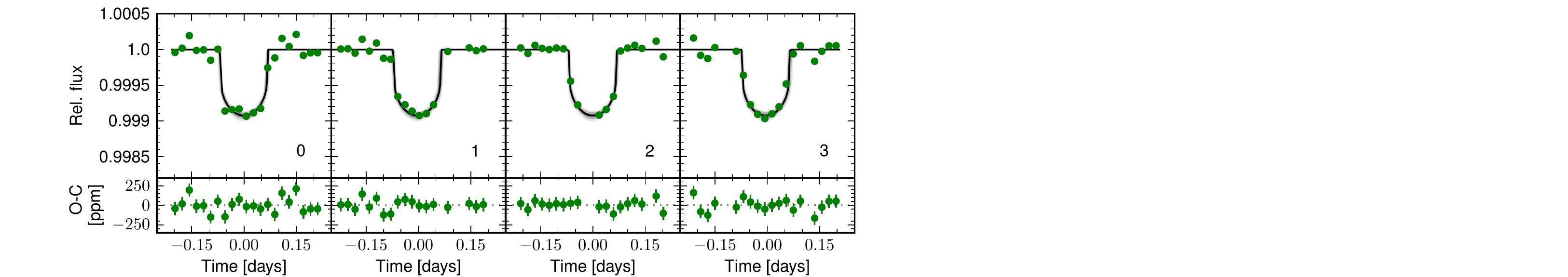}\hspace*{-7cm}\includegraphics[height=3.1cm]{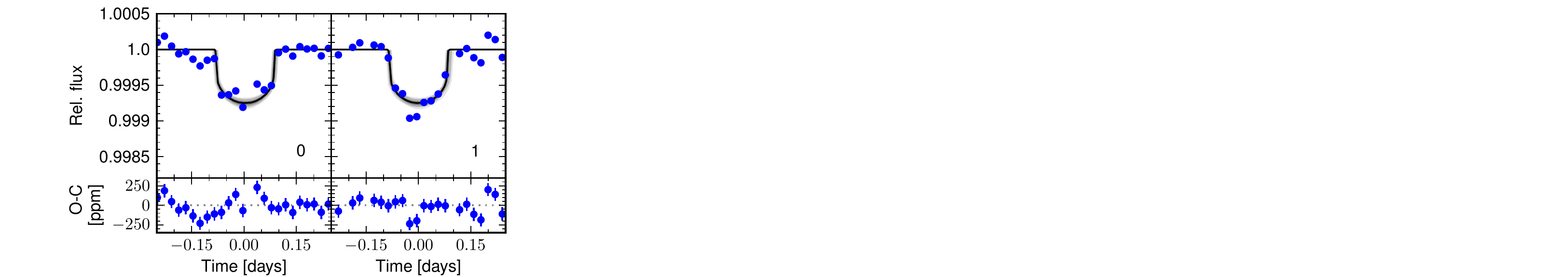}
\caption{As in Figure~\ref{figRV}, but for the \ktwo\ photometry of \epic\ (from top to bottom and left to right: planet~$b$, planet~$c$, and planet~$d$). Each transit is centered relative to the linear ephemeris.}
\label{figPH}
\end{figure*}

\end{appendix}

\end{document}